\documentclass[twocolumn,nofootinbib,aps,preprintnumbers,amsmath,amssymb]{revtex4-1}
\usepackage{amsmath}
\usepackage{graphicx}
\usepackage{dcolumn}
\usepackage{bm}
\usepackage{epsfig,color,xspace,multirow,xr,bbold}
\usepackage[all]{xy}
\usepackage{setspace}
\usepackage{url}
\usepackage[colorinlistoftodos]{todonotes}
\usepackage{threeparttable} 
\newcommand{\Ref}[1]{Ref.~\onlinecite{#1}}

\begin{document}

\title{
Electronic Spectra from TDDFT and Machine Learning in Chemical Space  
}

\author{Raghunathan Ramakrishnan}                                                         
\affiliation{Institute of Physical Chemistry and National Center for Computational Design and Discovery of Novel Materials, 
Department of Chemistry,                                                                  
University of Basel, Klingelbergstrasse 80, CH-4056 Basel,                                
Switzerland}                                                                              

\author{Mia Hartmann}                                                                 
\affiliation{Department of Chemistry and Biochemistry, California State University, Long Beach, 1250 Bellflower Boulevard, Long Beach, CA 90840, USA}

\author{Enrico Tapavicza}      
\email{Enrico.Tapavicza@csulb.edu}
\affiliation{Department of Chemistry and Biochemistry, California State University, Long Beach, 1250 Bellflower Boulevard, Long Beach, CA 90840, USA}
                                                                                       
\author{O. Anatole von Lilienfeld}                        
\email{anatole.vonlilienfeld@unibas.ch}                                 
\affiliation{Institute of Physical Chemistry and National Center for Computational Design and Discovery of Novel Materials,
Department of Chemistry,                                                                  
University of Basel, Klingelbergstrasse 80, CH-4056 Basel,                                
Switzerland}                                                                              
\affiliation{Argonne Leadership Computing Facility,                                   
Argonne National Laboratory, 9700 S. Cass Avenue, Lemont, IL 60439, USA}





\begin{abstract} 

Due to its favorable computational efficiency 
time-dependent (TD) density functional theory (DFT) 
enables the prediction of electronic spectra in a high-throughput manner across
chemical space. 
Its predictions, however, can be quite inaccurate. 
We resolve this issue with machine learning models trained on deviations of reference second-order approximate coupled-cluster singles and doubles (CC2) spectra from 
 TDDFT counterparts, or even from DFT gap.
We applied this approach to low-lying singlet-singlet vertical electronic 
spectra of over 20 thousand synthetically feasible small organic molecules 
with up to eight CONF atoms. 
The prediction errors decay monotonously as a function of training set size. 
For a training set of 10 thousand molecules, CC2 excitation energies
can be reproduced to within $\pm$0.1 eV for the remaining molecules. 
Analysis of our spectral database via chromophore counting suggests
that even higher accuracies can be achieved. 
Based on the evidence collected, we discuss open challenges associated with data-driven 
modeling of high-lying spectra, and transition intensities.
\end{abstract}

\maketitle


\section{Introduction}
Quantum mechanical rational compound design strategies~\cite{Beratan1996,fpdesign2014anatole} 
to model molecular valence electronic spectra holds great promise to narrow down the discovery of 
novel photonic, and optoelectronic devices. Potential applications include the fabrication of
low cost dye-sensitized solar cells \cite{gratzel2001photoelectrochemical}, organic light emitting diodes \cite{gross2000improving}, 
photosensitizers that are inert to environmental factors but useful in photodynamic therapy \cite{yogo2005highly}, and 
organic ultraviolet (UV) filters (aka sunscreens) in cosmetics~\cite{tan2014excited}. 
For any given compound, the relevant prediction accuracy can readily be attained with
an established excited state wavefunction method. 
Successful studies include the quantitative description
of solar cell materials \cite{pastore2010computational}, organic diodes \cite{han2011energy},
and even biologically relevant phenomena such as photo-induced dynamics of 
vitamins B2 \cite{wolf2008ultrafast}, and D \cite{Tapavicza2011}. 
For a robust forecast, depending on computational budget, one can also select 
a method according to the most appropriate cost-to-performance ratio from
the series of equations of motion (EOM) or linear 
response (LR) variants of the coupled cluster (CC)
theories CCS, CC2, CCSD, CC3 and CCSDT. 
These methods scale from $\mathcal{O}(N^4)$ to $\mathcal{O}(N^8)$,
where $N$ is the number of orbitals \cite{christiansen1995second}. 
When increasing size, or number of molecules, the next viable
compromise between accuracy and computational complexity 
is linear response time-dependent density functional theory (LR-TDDFT) within  
the adiabatic approximation~\cite{runge1984density,casida1995time}.
TDDFT, commonly based on local or semi-local exchange correlation (XC) 
functionals, has been shown to yield qualitatively inaccurate 
predictions whenever the valence excitations involve charge-transfer (CT)~\cite{dreuw2003long},
and the adiabatic approximation fails to accurately describe transitions with double excitation character~\cite{maitra2004double}.
Such qualitative failure of TDDFT, hard to anticipate 
without visual inspection of molecular orbitals involved in the transitions, 
dramatically reduces its usefulness for high-throughput screening of molecules
with interesting electronic spectra.
Application of CC methods for large scale computation is already prohibitive even when considering just electronic ground state properties of small sub-fractions of the known small molecule chemical universe, 
such as the GDB-17 with over 166 billion organic molecules
with no more than 17 atoms (not counting hydrogens). \cite{ruddigkeit2012enumeration}.

For combinatorially and computationally hard problems, such as
navigating chemical space in quest of an optimal electronic spectra~\cite{von2013first}, 
statistical inference from large volumes of data offers
an appealing alternative to the conventional strategies of investing
in ever more sophisticated approximations, faster hardware, or more
efficient programming. 
Statistical learning has already contributed to scientific progress 
in biology \cite{marx2013biology} or climate research \cite{mattmann2013computing}.
Inspired by the success of such efforts, several computational chemistry
studies have recently made use of supervised machine learning (ML) models to infer  
quantum mechanical properties of query molecules from those of 
a set of example molecules, computed {\it a priori}.
Effectively, this amounts to modeling expectation values 
calculated with approximate solutions to the electronic Schr\"odinger
equation, most notably the energy~\cite{rupp2012fast}.
By now, ML models have been shown
to reach the highly coveted quantum chemical accuracy
for many different ground-state molecular properties~\cite{montavon2013machine,hansen2013assessment,ramakrishnan201bigdata}. As such, also quantum mechanical expectation
values can be interpolated in chemical space~\cite{von2013first}.
Improvement of molecular models of chemical properties based on molecular similarity~\cite{Yaron1,Yaron2}
are also related to this approach.
These developments have also inspired studies on
transition state dividing surfaces~\cite{ML4Graeme2012},
orbital-free kinetic energy density functionals~\cite{ML4Kieron2012},
electronic properties of crystals~\cite{GrossMLCrystals2014},
transmission coefficients in nano-ribbon models~\cite{lopez2014modeling},
or densities of states in Anderson impurity models~\cite{arsenault2014machine}.
More recently a single kernel has been introduced for the simultaneous
 modeling of multiple electronic ground-state properties for training-sets comprised of up to 40,000 molecules~\cite{ramakrishnan2015many}.
Here, we report on our findings when trying to apply these ML methods to infer properties of molecules in their electronically excited states. 
More specifically, we discuss ML models which combine CC accuracy with DFT efficiency. 

\section{Methods}  
\label{sec:methods}

\subsection{$\Delta$-ML model of excited states properties}

In Ref.~\cite{ramakrishnan201bigdata}, some of us introduced the 
$\Delta$-ML {\it Ansatz} to estimate molecular 
ground-state properties from an expensive targetline theory, at the
computational cost of an inexpensive baseline theory (B). 
The ML model for the quantitative prediction of molecular electronic spectra 
is built in analogy, 
using ML models of the deviation of TDDFT excited state properties from CC2 reference numbers.
We approximate an electronic static property, $p_i$, corresponding to the $i^{\rm th}$ 
excited state of query molecule $q$ at CC2 level of theory 
as the sum of baseline prediction and a linear combination of
exponentially decaying functions in molecular similarity to training molecules $t$,
\begin{equation}
p_i^{\rm CC2}({\bf d}_q) \; \approx \; p_i^{\rm B} ( {\bf d}_q ) 
+ \sum_{t=1}^N c_{it} {\rm e}^{-|{\bf d}_q-{\bf d}_t|/\sigma},
\label{eq:krr}
\end{equation}
where $N$ is the number of molecules in training set, $\sigma$ is the kernel width,
and $|{\bf d}_q-{\bf d}_t|$ corresponds to the Manhattan ($L_1$) norm between 
molecular descriptors ${\bf d}$ ({\em vide infra}). 
A previous study~\cite{hansen2013assessment} benchmarked the performance of various norms in above equation
when directly modeling atomization energies (no baseline), 
and found the $L_1$ norm to yield lowest cross-validated errors.
The second term on the right side of Eq.~\ref{eq:krr} therefore models
exclusively the error in baseline method B's estimate of $p_i$ when compared to CC2 for query molecule $q$
\begin{eqnarray}
\Delta{p}_i^{\rm est}({\bf d}_q) =  \sum_{t=1}^N c_{it} {\rm e}^{-|{\bf d}_q-{\bf d}_t|/\sigma}.
\label{eq:del}
\end{eqnarray}
In this study we have investigate two excited state properties, $p_i$, namely excitation energy (with respect to the ground electronic state), $E_i$, and oscillator strength, $f_i$, 
for the lowest two ($i=1,\,2$) singlet electronic states.
Other excited states properties could have also been considered with this generic approach.
Due to their popularity we have selected for this study DFT and TDDFT as baseline B, and to CC2 as targetline. 
The CC2 method, with a triple-zeta basis set has been shown to predict experimental valence excitation energies with an MAE of 0.12 eV~\cite{send2011assessing}. 
This error decreases slightly to 0.10 eV, when CC2 is compared to the computationally more demanding method, CC3~\cite{kannar2014benchmarking}.
To serve as a reference method in this ML study we therefore consider CC2 to represent the optimal compromise between sufficient accuracy and acceptable computational cost.
To compare the impact of the baseline on the $\Delta$-ML strategy we have considered various DFT~\cite{HK,KS} baseline theories with increasing sophistication. 
However, any other combination of methods could have been chosen just as well.
Our simplest non-zero baseline for $p_i=E_i$, is the HOMO-LUMO gap of the ground-state computed using DFT-PBE0~\cite{PBE,PBE0,PBE01,adamo1999toward}. 
We also consider $p_i$ from LR-TDDFT~\cite{casida1995time,Furche2002} using the  hybrid functionals PBE0, and CAM-B3LYP~\cite{yanai2004new}.

In the following, we use matrix notations compatible with \Ref{ramakrishnan2015many}, and denote matrices by capital bold, 
and vectors by small bold cases.
Regression coefficients corresponding to training molecules, 
$\{c_{it}\}$, have
been obtained as solutions to
\begin{equation}
({\bf K}+\lambda {\bf I})\ {\bf c}_i = {\bf p}_i^{\rm CC2}-{\bf p}_i^{\rm B}=\Delta{\bf p}_i^{\rm ref},
\label{eq:l2}
\end{equation}
where ${\bf I}$ and ${\bf K}$ are the identity and kernel matrices, respectively,
the latter with elements $K_{st} = {\rm e}^{-|{\bf d}_s-{\bf d}_t|/\sigma}$. 
Note that in ML literature, the exponential kernel function is also denoted as Laplace kernel, owing to 
the fact that the exponential function, in certain coordinate systems, is a solution to Laplace's equation.
Eq.~\ref{eq:l2}
minimizes the $\lambda$-regularized ($\lambda$ quantifies the regularization strength)
least-squares error in estimations~\cite{arsenault2014machine}
\begin{equation}
\underset{{\bf c}_i}{{\rm minimize}} \quad || \Delta{\bf p}_i^{\rm ref} - \Delta{\bf p}_i^{\rm est}||_2^2 + \lambda {\bf c}_i^{\rm T}  {\bf K}  {\bf c}_i,
 \label{eq:fitness}
\end{equation}
where $||\cdot||_2$ stands for $L_2$ norm of a vector, $(\cdot)^{\rm T}$ denotes transpose operation,
and $\Delta{\bf p}_i^{\rm est}$ is defined in Eq.~(\ref{eq:del}).
Derivation of Eq.~(\ref{eq:l2}) from Eq.~(\ref{eq:fitness}) is presented as an Appendix.


\subsection{Cross-validation}

Overfitting of the kernel models to training molecules is typically avoided by optimizing the two hyperparameters ($\sigma$, $\lambda$) through five-fold cross validation (CV). 
In this procedure, $N$ training molecules are randomly distributed into 5 bins, each with $N$/5 molecules. 
Every bin is used once as a test (or validation) set, while the remaining four bins act as training sets. 
Hyperparameters are optimized such that they minimize the model's MAE for the test-bin. 
Here, we employed  Nelder-Mead's simplex method~\cite{nelder1965simplex} for the 2D optimization.
The cross validation procedure is the most time-consuming process in training the ML model, with
each evaluation of MAE of the test-bin requiring $\mathcal{O}(n^3)$ scaling matrix inversion operations,
where $n=4N/5$. In the present work $n$ is at most 8\,k. 
This results in roughly one CPU day of training for a fully converged ML model.

For larger training set sizes CVs become prohibitive, and one can employ the property-independent ``single-kernel'' Ansatz \cite{ramakrishnan2015many},
with optimal hyperparameters estimated exclusively from the structures of the training molecules. 
This approach assumes the training data to be devoid of outliers, and enforces $\lambda$ to be a fixed, property-independent scalar (typically set , or close, to zero). 
The width of the kernel function can be chosen according to some heuristic, for example such that the maximal value of the off-diagonal elements of ${\bf K}$
is $1/2$, which renders the kernel sufficiently global to have all training molecules contribute in the generation of the regression weights $\{c_{it}\}$.
For the exponential (aka Laplace) kernel, with $L_1$ distance metric, 
$K_{ij}=|{\bf d}_i-{\bf d}_j|/\sigma$, this constraint results in
$\sigma = {\rm max}\left\lbrace  |{\bf d}_i-{\bf d}_j|  \right\rbrace  / \log(2)$.
In \Ref{ramakrishnan2015many}, we have demonstrated a global kernel derived in this fashion to enable systematic 
reduction of out-of-sample prediction errors for thirteen molecular ground-state properties of 112\,k molecules, 
using up to 40\,k training molecules.
Here, in order to accelerate the CV procedure, we have made use of this heuristic as an initial guess. 
For the molecular datasets considered here these values in atomic units are typically $\sigma=1000$, and $\lambda=0$. 
After CV, globally optimal hyperparameters have been obtained by taking the median of the 5 folds. 
A median value is considered instead of mean, because
the median of a distribution is not influenced by extreme values, such as the hyperparameters that could
be found for a test bin with extreme outliers in structure or property. 
A final kernel with globally optimized hyperparameters is used for the prediction of the properties of out-of-sample molecules that are not part of training.

\subsection{Choice of molecular descriptor}
In order to assess the effect of the molecular representation on the ML model's performance, we report results based on 
two definitions of molecular representations, namely, the Coulomb-matrix (CM) with atom indices sorted by 
norm of rows in order to reach invariance with respect to permutation of identical nuclei \cite{rupp2012fast}, 
as well as a recently introduced more compact variant of the CM, called bag-of-bonds (BOB)~\cite{BobPaper}. 
The elements of the CM \cite{rupp2012fast} are defined as 
\begin{eqnarray}
M_{II} = 0.5 Z_I^{2.4},\qquad M_{IJ}=Z_I Z_J/R_{IJ},
\end{eqnarray}
where $I$, and $J$ are atomic indices, $R_{IJ}$ is the interatomic distance, 
and $Z$ is the atomic number. 
The off-diagonal elements of a CM uniquely represent the geometry and atomic composition of a molecule~\cite{von2015fourier}, 
while the diagonal elements 
provide a simple exponential fit to the negative of the potential energy
of the neutral atoms. 
As such, the diagonal is similar to the
total potential energy of a neutral atom within Thomas-Fermi theory, 
$E_{\rm TF}=-0.77 Z^{7/3}$, or its
modifications with a $Z$-dependent prefactor in the range 0.4--0.7~\cite{parr1989density}. 
It is sufficient to consider only the lower or upper triangle of the CM. In order to enable
comparisons between two molecules with different number of atoms, the CM matrix of the
smaller molecule is padded with zero elements.

BOB is a labeled set of off-diagonal CM elements which enables
the comparison of pairwise distance between any given combination of two atom types. 
For instance, for H$_2$O, BOB is the set of two sorted row vectors,  
$\left\lbrace  \left[ M_{{\rm HO}}, M_{{\rm HO}} \right],  \left[M_{{\rm HH}}\right]  \right\rbrace$,
with elements corresponding to the CM entries.
Due to the pairwise partitioning, however, any two homometric molecules with identical stoichiometry
will yield a zero descriptor difference according to  BOB~\cite{von2015fourier}.
As such, BOB does not uniquely represent molecules.  
The CM, by contrast, is able to uniquely encode any molecule, up to its enantiomers. 
The molecular dataset considered in this study, however, is devoid of homometric molecules.
In addition to the aforementioned sorted CM matrix, BOB has also been tested 
since it has been shown to yield slightly better accuracy for the prediction of molecular atomization energies \cite{BobPaper}. 
In general, 
we have found that for large $N$, both CM and BOB converge towards similar prediction accuracy for energy-related properties. 
For smaller training sets, however, BOB typically exhibits a more substantial advantage.

\begin{figure*}[hpt]
\centering      
\includegraphics[width=17.8cm, angle=0.0, scale=1]{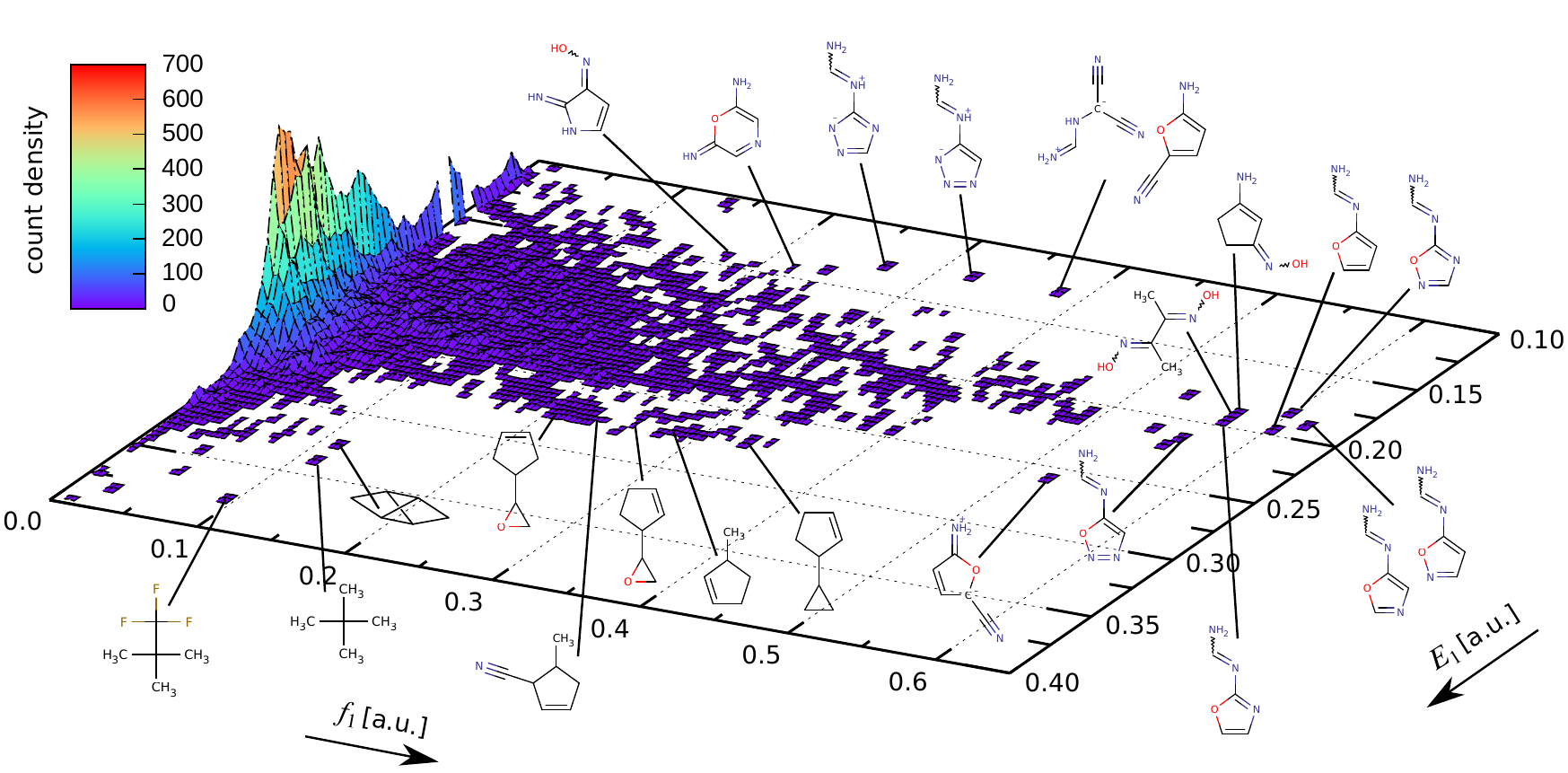}                              
\caption{Joined distributions of oscillator strength $f_1$, and transition energy $E_1$ for the first electronic excited singlet state of the 22\,k organic molecules. 
All values correspond to the RICC2/def2TZVP level of theory.
For select $E_1$ values, representative molecules with large $f_1$ are shown as insets.} 
\label{fig:jointdist}
\end{figure*}

\subsection{Excited states data} 
\label{sec:comp}
We have relied on the recently published molecular quantum chemistry database with relaxed geometries
computed using the DFT B3LYP with basis set 6-31G(2df,p)~\cite{ramakrishnan2014quantum}. 
This data set corresponds to the smallest 133,885 (134\,k) organic molecules with up to 9 CONF atoms 
out of the list of 166\,B synthetically feasible organic molecules, called GDB-17 database, and published by 
Reymond and co-workers~\cite{ruddigkeit2012enumeration}.
For this study, 
we have eliminated 3,054 molecules from the 134\,k dataset 
due to high steric strain in the B3LYP/6-31G(2df,p) geometries~\cite{ramakrishnan2014quantum},
and we further have limited ourselves to those molecules 21,800 molecules with only up to 8 CONF atoms. 
For these molecules we have performed single point calculations using the program TURBOMOLE \cite{TURBOMOLE}
to compute the ground (S$_0$)~\cite{Haeser1989}, and the lowest two vertical
electronic excited states (S$_1$ and S$_2$) of singlet spin-symmetry. 
We also performed calculations at the LR-TDDFT~\cite{Furche2002} level employing the 
hybrid XC functional PBE0 \cite{PBE0,adamo1999toward} with def2SVP basis set \cite{weigend2005balanced}, 
and at the resolution-of-identity approximate coupled cluster with singles and doubles substitution (RI-CC2)~\cite{hattig2000cc2} 
level with def2TZVP basis set \cite{weigend2005balanced}.
Using the larger basis set, we also performed LR-TDDFT  calculations
with PBE0, and CAM-B3LYP~\cite{yanai2004new} functionals, the latter 
using the program Gaussian09 \cite{ftssrcsbmpetal2009}. 

All calculations were performed with C$_1$ symmetry and in
DFT calculations default integral grids were employed to compute the XC energy contributions.
For 7 molecules (most of them highly symmetric, e.g.~cubane) 
the RI-CC2 calculations did not converged the first excited state wavefunction. 
For 7 other molecules (with multiple CO groups, e.g.~2,3-dioxobutanedial) 
emmission has been found, i.e.~negative lowest transition energy, presumably arising from orbital relaxation. 
For the purpose of this study, we have removed these exotic 14 molecules. 
The lowest two singlet transition energies, as well as corresponding oscillator strengths in length-representation,
 have been used for the remaining 21,786 molecules, to which we refer in the following as the set of 22\,k molecules.
All indices of the 22\,k GDB-8 molecules 
along with corresponding TDDFT, and CC2 excitation energies are given as supplementary material in 
{\tt gdb8\_22k\_elec\_spec.txt}. 
The indices enable retrieval of geometries from
the 134\,k GDB-9 dataset~\cite{ramakrishnan2014quantum}.

\section{Results and discussion}
\label{sec:results}
\subsection{Excitation energies and oscillator strengths for 22\,k organic molecules}

The smoothened distribution of CC2 predicted S$_0\rightarrow$S$_1$ transition energies $E_1$ and corresponding oscillator strengths $f_1$ features in Figure~\ref{fig:jointdist} for all 22\,k molecules. 
This 2D count density has been computed via kernel-density estimation \cite{silverman1986density,botev2010kernel}.
The first excitation energy distribution is bimodal (see also Fig.~\ref{fig:density} for the 1D projection), 
corresponding to one Gaussian centered 
at 0.18 a.u.~with small variance, and another centered near 0.26 a.u.~with significantly larger variance (the shoulder possibly implying
two peaks, rather than one broad peak). 
Collectively, the 22\,k molecules span the spectral range of UV-B and UV-C,
with few molecules in the UV-A region ($>$ 300 nm or $<$ 0.15 a.u.). 
The lack of transitions in the visible region is consistent with the fact that small organic molecules typically 
exhibit an energy gap of $>$ 5 eV between highest and lowest occupied molecular orbital, HOMO and LUMO respectively. 
Not surprisingly, when proceeding from low to high transition energy regions one notices that molecules gradually turn from being aromatic, 
or highly unsaturated, into increasingly saturated structures.
The oscillator strength ($f_1$), by contrast, exhibits an exponentially decaying distribution,
with the largest fraction of compounds in the 22\,k set having negligible or zero values.
A small minority of molecules, however, have significant$f_1$-magnitude, 
implying potential usefulness of these molecules as components in 
metal-free organic sensitizers \cite{tseng2014design}. 
About a dozen molecules, highlighted in Figure~\ref{fig:jointdist}, 
display $f_1 > 0.5$, resulting in light harvesting efficiencies~\cite{memming2008semiconductor} larger
than $100 \times (1-10^{-0.5}) \approx $ 68\%. 
These molecules contain ketoxime, R(R$^\prime$)C=NOH,
or amidine, R-C(NH$_2$)=NR$^\prime$, chromophores.
They all exhibit push-pull type conjugation of $\pi$-bonds, with electron-donating, and
electron-withdrawing groups on opposite ends, resulting in highly polarized electron densities. 
However, also the symmetric molecule 
(point group C$_{2\rm h}$), dimethylglyoxime, 
a chelating agent commonly used in gravimetric analysis of nickel, 
has a large oscillator strength for its first excitation with 
$f_1$ = 0.56 at $E_1$ = 0.2 a.u.  

The effect of level of theory is shown for TDPBE0 and CC2 predictions of $E_1$ 
and $E_2$ in the top panel of Figure~\ref{fig:density}. 
For both states,  TDDFT leads to a depletion in count densities at
$\thickapprox$ 7 eV when compared to the CC2 distribution, 
compensated by overestimated densities in low and high energy regions. 
In the following we will discuss how to mitigate this depletion 
using ML models. 

\subsection{ML models of excitation energies}  
\begin{figure}[hpt]
\centering      
\includegraphics[width=8.5cm, angle=0.0, scale=1]{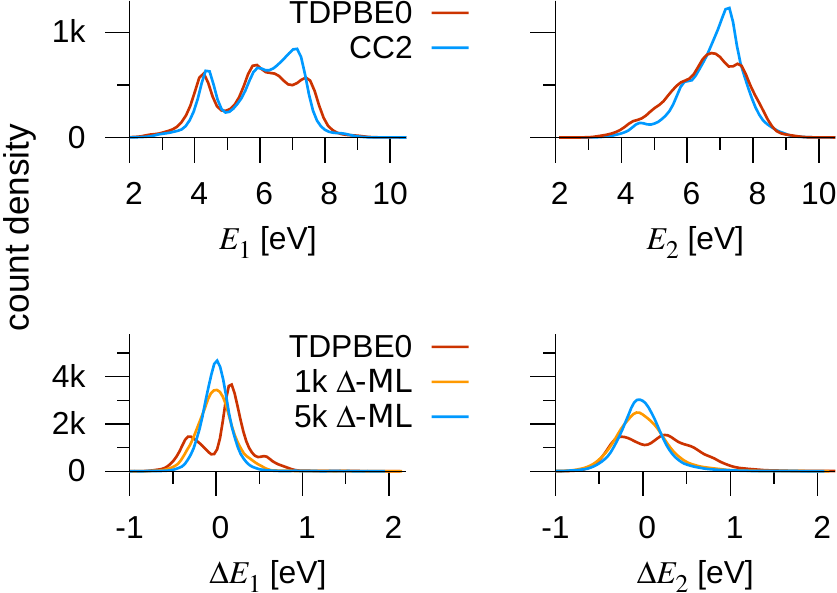}                              
\caption{
Distributions, smoothened by 1D kernel density estimation as implemented in GNUPLOT~\cite{williams2010gnuplot}, of spectral properties and predicted errors. 
Top: Densities of first, and second singlet transition energies 
($E_1$, and $E_2$, respectively, in eV) of 17\,k organic molecules with up to eight CONF atoms, at the CC2, and
TDPBE0 levels of theory. 
Bottom: Error distribution for $E_1$ (left) and $E_2$ (right) with respect to CC2. 
Errors are given for TDPBE0 and $\Delta$-ML models based on 1\,k, and 5\,k training molecules with TDPBE0 baseline.} 
\label{fig:density}
\end{figure}

Despite the obvious differences in prediction in the top panel of Fig.~\ref{fig:density},
the $\Delta$-ML model of Eq.~(\ref{eq:krr}) captures the necessary correction. 
This is illustrated by the signed error distributions (with respect to CC2)
in the bottom panel of Figure~\ref{fig:density}, for both excitation energies.
Distributions are shown for $\Delta$-ML models trained on molecular sub-sets 
containing either $N =$ 1\,k or $N = $ 5\,k molecules, drawn at random from the 22\,k data set. 
All ML results discussed in this paper, including these 
distributions, correspond to out-of-sample predictions for the remaining  (22\,k - $N$) molecules. 
For comparison, the TDDFT deviation from CC2 is also shown in the bottom panel 
of Fig.~\ref{fig:density} for both
transition energies, resulting in a bimodal distribution which suggests that systematic errors are present. 
These errors are can be either due to PBE0 kernel, or smaller basis-set, or both. 
The ML errors, by contrast, are normally distributed around zero,
with increasing and decreasing height and width, respectively, as
one increases the training set from 1\,k to 5\,k. 
This implies that the $\Delta$-ML model is properly accounting for the systematic
errors in the TDDFT predictions, replacing them by a normal error distribution. 
Mean absolute errors (MAEs) of the TDDFT predictions amount to 
of 0.27, and 0.37 eV, for $E_1$, and $E_2$, respectively. 
These MAEs are reduced to 0.16, and 0.23 eV for the 1\,k ML models, 
and to 0.13 and 0.20 eV for the 5\,k ML models. 
We have also investigated the effect of the other aforementioned descriptor in the ML model, BOB. 
BOB results in ML prediction errors of 0.13/0.20 and 0.09/0.16 eV for $E_1$/$E_2$, using models trained on
1\,k and 5\,k training sets, respectively--- slightly better than the corresponding CM predictions.

\begin{figure*}[hpt]
\centering      
\includegraphics[width=17.8cm, angle=0.0, scale=1]{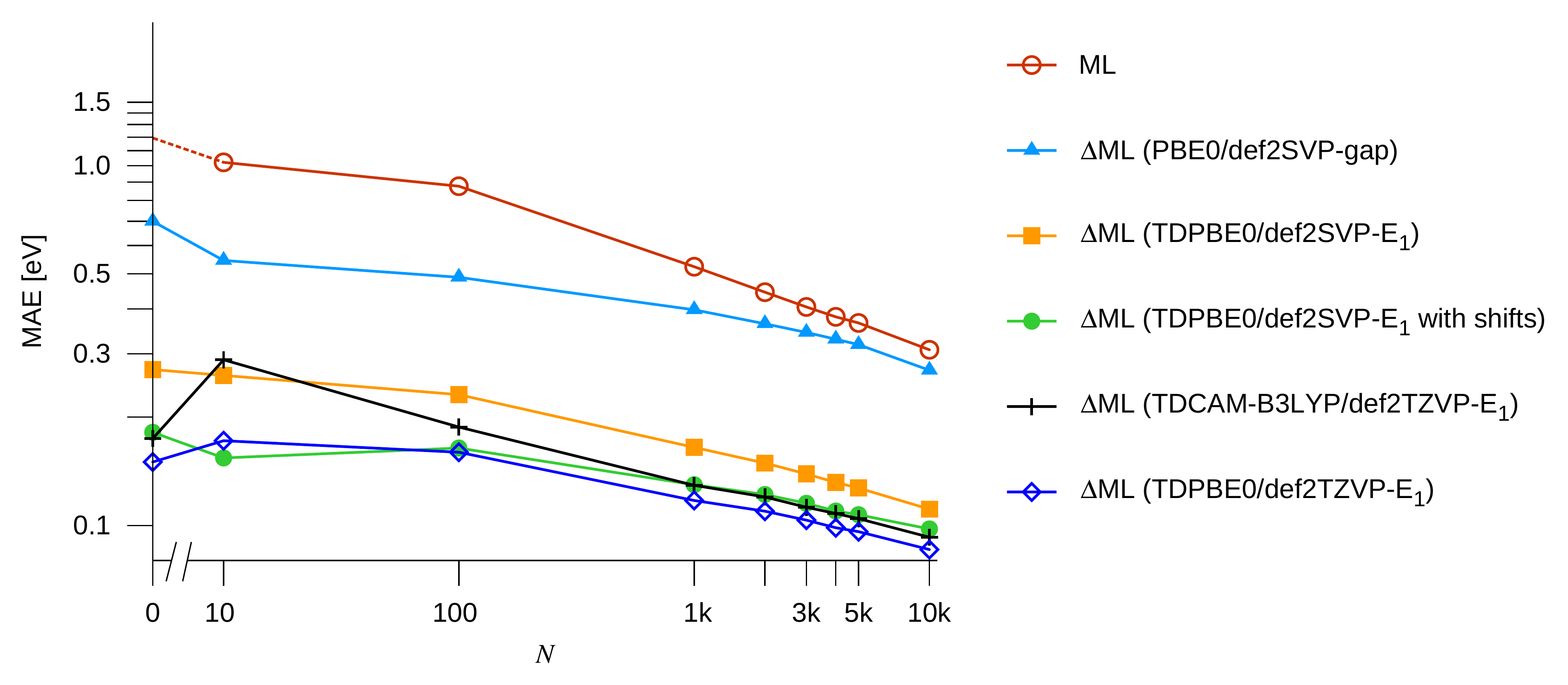}                              
\caption{
Systematic improvement of ML models of singlet-singlet transition  energies ($E_1$). 
Mean absolute error (MAE in eV) with respect to reference CC2/def2TZVP values is shown as a function of 
trainingset size ($N$) for 22\,k$-N$ out-of-sample predictions. 
Various baseline methods are shown.
The value of the baseline-free ML model (red) at $N=0$ corresponds to the CC2 standard deviation in the 22\,k test set.
Other baselines include HOMO-LUMO gap (blue), 
TDPBE0 $E_1$ without (yellow), and with (green) 
bivariate systematic shift corrections which explicitly account for $\sigma$ and $\pi$ chromophores.
Also included are def2TZVP baseline results for TDCAM-B3LYP $E_1$ (black) and TDPBE0 $E_1$ (blue). 
Baseline errors at $N=0$ correspond to standard deviations, obtained after subtraction of an average shift 
with respect to the CC2-targetline.
} 
\label{fig:ml}
\end{figure*}

In order to investigate in a systematic fashion the performance of the $\Delta$-ML model, 
we have calculated out-of-sample MAEs of $E_1$ predictions for various baseline methods. 
In Figure \ref{fig:ml} the resulting MAEs are shown as a function of training set size $N$
for $N$ = 0 (i.e. the error of the baseline method), 10, 100, 1\,k,
2\,k, 3\,k, 4\,k, 5\,k, and 10\,k. 
More specifically, zero baseline results correspond to setting $E_1^{\rm B}$ to zero in Eq.~(\ref{eq:krr}).
We also used the PBE0 HOMO-LUMO gap as a baseline, as well as TDPBE0 and TDCAM-B3LYP. 
As one would expect, the predictive accuracy improves with increasing level of sophistication of the baseline:
The zero, gap, and TD baseline with def2SVP basis set yield 0.4, 0.3, and 0.13 eV,
respectively, for the most accurate model trained on $N = $ 10\,k molecules.
Increasing the basis set from def2SVP to def2TZVP improves
PBE0's baseline value, eventually resulting in a very small MAE of 0.08 eV
for 10\,k $\Delta$-ML. 
These observations are in line with previous benchmark 
calculations~\cite{jacquemin2009extensive} which concluded 
TDCAM-B3LYP is somewhat inferior to TDPBE0 for the prediction of
singlet-to-singlet excitation energies of small molecules. 
Overall, it is encouraging that all models, no matter which 
baseline, converge towards the same learning rate, i.e.~slope on the log-log
scale of error versus training set size. 
As such, the baseline merely leads to a difference in off-sets
--- which could also be compensated for by adding more training data. 
Due to the immense size of chemical space~\cite{von2013first}, the addition of more molecules can easily be envisioned.
For $\Delta$-ML models of $E_2$, similar curves can be obtained,
albeit slightly off-set yielding less accurate predictive power. 

\subsection{Inclusion of systematic shifts}

\begin{figure}[hpt]
\centering      
\includegraphics[width=8.5cm, angle=0.0, scale=1]{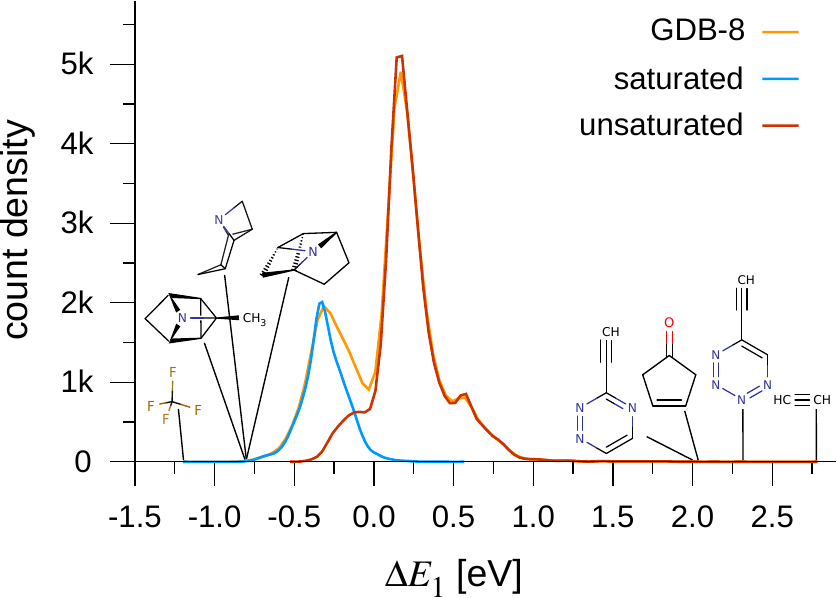}                              
\caption{
Bivariate  error distribution of the TDPBE0/def2SVP  
lowest singlet-singlet transition energies ($E_1$ in eV) of
22\,k organic molecules with up to eight CONF atoms (yellow). 
Partitioned error distributions over saturated
(blue) and unsaturated (red) molecules are shown as well. 
The molecular structures correspond to extreme outliers for TDPBE0/def2SVP.
} 
\label{fig:biv}
\end{figure}

It is not obvious to us that there is a single reason for TDPBE0/def2SVP's substantial underestimation of first and second
transition energies near 7 eV, see Figure~\ref{fig:density}.
A simple pattern, however, emerges after splitting the 22\,k set into
saturated and unsaturated molecules, i.e.~into two sets containing either $\pi$- or $\sigma$-chromophores. 
The corresponding signed error densities for the two sets are well separated, 
as shown for $E_1$ in Figure~\ref{fig:biv}.
They are centered around -0.31, and +0.19 eV for the saturated $\sigma$
and unsaturated $\pi$-chromophores, respectively.
The systematic underestimation of  TDPBE0-based $E_1$ of
$\pi$-type excitations ($\pi \rightarrow \pi^*$ or $n \rightarrow \pi^*$) is a well-known
issue of approximate XC functionals when it comes to the description of CT-type 
excitations~\cite{dreuw2003long}, i.e. transitions with small overlap between donor and acceptor orbital overlap \cite{Peach2008}. 
Our results are consistent with this finding, strengthening the indications that the 
underestimation of $E_1$ is universal for all $\pi$-type excitations. 
Furthermore, the other distribution in Figure~\ref{fig:biv}
clearly shows a systematic overestimation of TDPBE0-based $E_1$ of $\sigma$-type excitations 
($\sigma \rightarrow \sigma^*$ or $n \rightarrow \sigma^*$).
This systematic blue shift of TDPBE0 $E_1$ is, at least partly, due to
the finiteness of the relatively small basis set (def2SVP) used. 
This reasoning is in line with the variational principle: The difference
between the lowest two eigenvalues of the molecular Hamiltonian is always larger 
when represented in a small basis set than when compared to the complete basis set limit.
For instance,  using literature values \cite{johnson2013nist} of
the HOMO-LUMO gap of the water molecule, we note the PBE0 value
with the minimal basis set, STO-3G, to be 13.3 eV, overestimating more 
converged basis set PBE0 results by roughly 4.6 eV. 

The degree of saturation can easily be detected beforehand using SMILES strings. 
We can therefore readily exploit this knowledge by subtraction of the distribution's centered value -0.31, and +0.19 eV from the baseline number
for saturated and unsaturated chromophores, respectively.
The resulting TDPBE0 $\Delta$-ML model in Eq.~(\ref{eq:krr}) improves indeed:
The out-of-sample MAE decreases at a lower off-set with training set size, 
as shown in Fig.~(\ref{fig:ml}), yet at similar learning rates as the other models.  
For the 10\,k model of the $E_1$ transition energy the MAE
is found to decrease from 0.13 eV to 0.1 eV. 
It is intersting to note that the performance of the TDCAM-B3LYP/def2TZVP level is 
virtually identical with the shifted TDPBE0/def2SVP result ($N = 0$), as well as for larger $N$ values. 
For smaller training sets ($N$ = 10 or 100), the shifted TDPBE0/def2SVP $\Delta$-ML model even outperforms the corresponding
TDCAM-B3LYP/def2TZVP variant. 

\subsection{DFT and ML model outliers}

It is always interesting to consider the worst predictions of a model.
The average errors discussed so far neither imply better ML predictions for DFT outliers nor do they quantify the ML outliers. 
Here, we briefly discuss the accuracy of predicted $E_1$ for the 10 most extreme outliers among all out-of-sample
molecules, i.e.~all molecules that were not part of the training sets for the 1\,k, and 5\,k $\Delta$-ML models. 
Table~\ref{tab:outlier} lists SMILES strings of corresponding molecules, model prediction errors, and CC2 numbers for comparison. 
The 10 outliers are sorted by their TDPBE0, 1\,k, or 5\,k ML model deviation. 
As also already indicated in Figure~\ref{fig:biv}, the worst DFT outliers correspond to unsaturated molecules. 
This observation holds true for the 10 most extreme DFT outliers in Table~\ref{tab:outlier}, deviating by up to 2.15 eV from CC2. 
These molecules could be of interest as benchmarks for developing improved DFT kernels for TDDFT calculations.
The numbers in Table~\ref{tab:outlier} show that for all outliers, the 5\,k ML model yields better performance than DFT,
while the 1\,k ML model improves all predictions but the one for the worst, namely cyclopenta-1-en-4-on ({\tt O=C1CC=CC1}). 
This molecules is also shown in Fig.~\ref{fig:biv}. 
Note that other outliers shown in that figure have been part of the training set and therefore do not feature in Table~\ref{tab:outlier}.
The finding that the ML models also improve on the baseline method's outliers agrees with conclusions drawn in a previous findings
where we applied the $\Delta$-ML Ansatz to model DFT-level enthalpies of atomization for 
the 134\,k dataset, and where we found that for the most extreme outlier the baseline model's error reduced
systematically with the training set size of the augmenting ML model~\cite{ramakrishnan201bigdata}. 

When considering the 10 most extreme outliers of the ML models in Table~\ref{tab:outlier},
neither order nor identity of the DFT outliers is conserved.
Among the top 10 outliers of the 10 k\,model, for example, there is even a saturated molecule from the opposite (blue) end
of the error distribution in Fig.~\ref{fig:biv}: tetra-fluoro-methane CF$_4$, with an underestimating deviation -1.27 eV.


\begin{table}[!htp]                                                                       
\begin{threeparttable}[b]  
\caption{ 
10 most extreme outliers for TDPBE0/def2SVP and $\Delta$-ML models.
Largest deviations of predicted lowest singlet-singlet transition energy ($E_1$) from corresponding CC2/def2TZVP value. 
All values in eV.
        }                                                                                 
\label{tab:outlier}                                                                           
\begin{tabular}{l   |c|c|c||c }
\hline 
\hline 
\multicolumn{ 1}{l}{Molecule} &                                                           
\multicolumn{ 1}{c}{TDPBE0} &                                                             
\multicolumn{ 1}{c}{1k $\Delta$-ML} &
\multicolumn{ 1}{c}{5k $\Delta$-ML} &
\multicolumn{ 1}{c}{CC2} \\[0.5ex]  
\hline 

top DFT outliers &&&&\\
      {\tt  FC1=COC=NC1=O }     &          1.63      &     1.41  &         1.40      &     5.85    \\
      {\tt  CC1=COC=CC1=O }     &          1.64      &     1.22  &         1.04      &     5.69    \\
      {\tt  CC1=CC(=O)C=NO1 }   &          1.66      &     1.08  &         0.88      &     5.23    \\
      {\tt  CC1=C(NN=N1)C=O }   &          1.73      &     1.55  &         1.51      &     5.57    \\
      {\tt  O=CC1=CN=CN=C1 }    &          1.81      &     1.50  &         1.42      &     5.38    \\
      {\tt  CC1=C(C)CC(=O)C1 }  &          1.82      &     1.62  &         1.56      &     6.12    \\
      {\tt  CN1C=C(C=O)C=N1 }   &          1.84      &     1.62  &         1.35      &     5.82    \\
      {\tt  CC1=C(C=O)N=NO1 }   &          1.92      &     1.52  &         1.74      &     5.82    \\
      {\tt  C\#CC1=NC=CN=N1 }    &          1.99      &     1.43  &         1.76      &     5.02    \\
      {\tt  O=C1CC=CC1 }        &          2.13      &     2.15  &         1.95      &     6.44    \\\hline
top 1k $\Delta$-ML outliers &&&&\\
  {\tt  O=N(=O)C1=NC=CO1 } &               1.53   &   1.33     &      1.41    &       5.31        \\
  {\tt  FC1=COC=NC1=O }    &               1.63   &   1.41     &      1.40    &       5.85        \\
  {\tt  C\#CC1=NC=CN=N1 }   &               1.99   &   1.43     &      1.76    &       5.02        \\
  {\tt  CC(=O)C1=CC=NN1 }  &               1.62   &   1.46     &      0.91    &       5.55        \\
  {\tt  O=CC1=CN=CN=C1 }   &               1.81   &   1.50     &      1.42    &       5.38        \\
  {\tt  CC1=C(C=O)N=NO1 }  &               1.92   &   1.52     &      1.74    &       5.82        \\
  {\tt  CC1=C(NN=N1)C=O }  &               1.73   &   1.55     &      1.51    &       5.57        \\
  {\tt  CC1=C(C)CC(=O)C1 } &               1.82   &   1.62     &      1.56    &       6.12        \\
  {\tt  CN1C=C(C=O)C=N1 }  &               1.84   &   1.62     &      1.35    &       5.82        \\
  {\tt  O=C1CC=CC1 }       &               2.13   &   2.15     &      1.95    &       6.44        \\\hline
top 5k $\Delta$-ML outliers &&&&\\
      {\tt  O=N(=O)C1=NC=CO1 } &           1.53    &   1.33    &     1.41   &            5.31   \\
      {\tt  FC(F)(F)F }        &          -1.20    &  -1.27    &    -1.42   &           13.98   \\
      {\tt  O=CC1=CN=CN=C1 }   &           1.81    &   1.50    &     1.42   &            5.38   \\
      {\tt  O=CC1=NC=CC=C1 }   &           1.58    &   1.31    &     1.46   &            5.09   \\
      {\tt  CC1=C(NN=N1)C=O }  &           1.73    &   1.55    &     1.51   &            5.57   \\
      {\tt  OC1=NOC(C=O)=C1 }  &           1.59    &   1.32    &     1.54   &            5.18   \\
      {\tt  CC1=C(C)CC(=O)C1 } &           1.82    &   1.62    &     1.56   &            6.12   \\
      {\tt  CC1=C(C=O)N=NO1 }  &           1.92    &   1.52    &     1.74   &            5.82   \\
      {\tt  C\#CC1=NC=CN=N1 }   &           1.99    &   1.43    &     1.76   &            5.02   \\
      {\tt  O=C1CC=CC1 }       &           2.13    &   2.15    &     1.95   &            6.44   \\\hline

\hline 
\hline 
\end{tabular}                                                                             
\end{threeparttable}                                                                      
\end{table}

\subsection{ML models of oscillator strengths}
We have also investigated the applicability of the $\Delta$-ML Ansatz to model
oscillator strengths, $f_1$ and $f_2$ for ${\rm S}_0 \rightarrow {\rm S}_1$ and ${\rm S}_0 \rightarrow {\rm S}_2$ transitions, respectively.  
While the $\Delta$-ML models of excitation energies can be systematically improved through mere addition of training data,
corresponding models for $f_1$ or $f_2$ do not become more accurate with increasing training set size. 
TDCAM-B3LYP has been show to yield oscillator strengths with minimal deviations with respect to correlation TD methods ~\cite{caricato2010oscillator}. 
For our 22\,k dataset, TDCAM-B3LYP/def2TZVP yields an MAE of 0.0101 a.u., 
compared to CC2/def2TZVP. 
This deviation is reduced to only 0.0100, and 0.0099 a.u. when augmenting the CAM-B3LYP numbers with $\Delta$-ML models trained on 1\,k, and 5\,k molecules, respectively. 
Also changing the descriptor from CM to BOB did not improve the state of affairs.

The $\Delta$-ML model approach might fail for several reasons.
For one, $f_i$ is a rather complex property which requires knowledge of a certain combination of two wave-functions, 
\begin{eqnarray}
f_i \varpropto |\langle 0|\hat{ \mu}  |i\rangle|^2 E_i.
\end{eqnarray}
This could imply the need for substantially larger training sets in order to obtain satisfying learning curves.
Another explanation might be that the training problem is ill posed. 
In fact, TDDFT often yields a different ordering of states than CC2, 
implying that the baseline property corresponds to a different matrix element than the targetline property.
This, in turn, will also result in substantially less efficient ML training scenarios. 
However, this reasoning, while appealing to explain the failure of a
$\Delta_{\rm TDDFT}^{\rm CC2}$-ML model, does not satisfyingly 
explain why also a direct ML model with zero baseline shows insignificant
prediction improvement with increasing training set size. 
Finally we remark that also previously we have seen significantly
less impressive learning rates for other electronic integrals, 
e.g.~the norm of the molecular dipole moment in organic molecules \cite{ramakrishnan2015many}.

\section{Conclusions}
In summary, we have applied the $\Delta$-ML approach, previously introduced
to accurately model molecular ground state properties, to the data-driven modeling of electronic excitation energies. 
We have computed the low-lying valence electronic spectra for a modest chemical universe of 22\,k organic molecules, 
made up from up to 8 CONF atoms, at the level of TDDFT (using PBE0, and CAM-B3LYP), and CC2. 
We have presented numerical evidence that large basis set CC2-level valence excitation energies can be estimated 
at the speed of small basis set TDPBE0 through statistical inference of the difference, 
derived from training on a fraction of this database.

Analysis of the data-sets, based on kernel density estimates, suggests small basis set TDPBE0 level of theory
to over-, and under-estimate the lowest two transition energies for organic molecules
with $\sigma$-, and $\pi$-chromophores, respectively. 
This behavior results in well separated bivariate error distribution. 
Accounting for these systematic shifts enables further improvement of the $\Delta$-ML models.
From a methodological point of view, this procedure allows to readily integrate
expert knowledge of error distributions in the ML model, resulting in improved predictions.
For an automated estimation of systematic shifts arising from multivariate property distributions, one
can adapt clustering protocols based on kernel density estimates. 
Such clustering has been done previously in the context of analyzing Monte Carlo 
trajectories~\cite{christov2011correlated}, collective variables in molecular 
dynamics~\cite{wang2012entangled,gasparotto2014recognizing}, or even to quantify the
contribution of an MO to total electronic energy~\cite{melnichuk2012relaxed}.

The numerical evidence for the modeling of excitation energies suggests that 
severe flaws in TDDFT based predictions can easily be rectified through statistical learning, 
irrespective of their origin such as possible incorrect state ordering, basis set incompleteness, 
inherent limitations of adiabatic TDDFT for states with doubly excited, or CT character. 

The poor performance of ML models for predicting oscillator strengths 
warrants future investigations. 
The database of excited states properties for 22\,k organic molecules
(see Supporting Information) might also be useful for 
benchmarking the performance of other approximations and models, as well as to facilitate 
the identification of  potential, hitherto unknown, chromophore-auxochrome relationships. 
Eventually, our study might aid the computational design of 
functional molecular components with desirable photochemical properties.

\section{Acknowledgement}
OAvL acknowledges funding from the Swiss National Science Foundation (No. PP00P2\_138932). 
ET acknowledges start-up funds from California State University Long Beach. 
Some calculations were performed at sciCORE (http://scicore.unibas.ch/) scientific computing core facility at 
University of Basel. This research used resources of the Argonne Leadership Computing Facility at Argonne 
National Laboratory, which is supported by the Office of Science 
of the U.S. DOE under contract DE-AC02-06CH11357. 

\section*{Appendix: Derivation of Eq.~\ref{eq:l2} in matrix notation}
We derive the linear system of equations in Eq.~\ref{eq:l2} by employing the 
regularized least squares error measure, Eq.~\ref{eq:fitness}. 
Let us denote the reference property values of training molecules as the column vector ${\bf x} = {\bf p}^{\rm ref}$. 
The Kernel-Ridge-Regression Ansatz for the estimated property values of 
training molecules is ${\bf p}^{\rm est}={\bf K}{\bf c}$. 
The $L_2$-norm of the residual vector, penalized by regularization of fit coefficients, is the Lagrangian 
\begin{eqnarray}
\mathcal{L}   & = & ||{\bf p}^{\rm ref}-{\bf p}^{\rm est}||_2^2 + \lambda {\bf c}^{\rm T}{\bf K}{\bf c} \nonumber \\
             & = & \left( {\bf x}-{\bf K}{\bf c} \right)^{\rm T} \left( {\bf x}-{\bf K}{\bf c} \right) +  \lambda {\bf c}^{\rm T}{\bf K}{\bf c}   \nonumber \\ 
            & = & {\bf x}^{\rm T} {\bf x}-{\bf x}^{\rm T} {\bf K}{\bf c}-({\bf K}{\bf c})^{\rm T} {\bf x}+ ({\bf K}{\bf c})^{\rm T} {\bf K}{\bf c}+  \lambda {\bf c}^{\rm T}{\bf K}{\bf c},\nonumber \\
           & &  
\end{eqnarray}
 where $(\cdot)^{\rm T}$ denotes transpose operation.
To minimize the Lagrangian, we equate its derivative with respect to the regression coefficients vector, 
${\bf c}$, to zero
\begin{eqnarray}
\frac{d}{d {\bf c}} \mathcal{L} = 
    -{\bf x}^{\rm T} {\bf K}-{\bf K}{\bf x} + {\bf K}{\bf K}{\bf c} +{\bf c}^{\rm T} {\bf K}{\bf K} +  &   &   \nonumber \\
 \lambda {\bf K}{\bf c} +   \lambda {\bf c}^{\rm T} {\bf K} & = & 0.
\label{eq:final}
\end{eqnarray}
Here we have used 
the fact that the kernel matrix ${\bf K}$ is symmetric, i.e., ${\bf K}^{\rm T} = {\bf K}$
along with the matrix calculus identity, $\left( d/d{\bf c}   \right){\bf c}^{\rm T}= {\bf I}$, where ${\bf c}$ is a column vector
and ${\bf c}^{\rm T}$ is a row vector. Grouping by row and column vectors yields
\begin{eqnarray}
\left(  {\bf K}{\bf K}{\bf c} +  \lambda {\bf K}{\bf c} - {\bf K}{\bf x} \right) + 
\left(  {\bf K}{\bf K}{\bf c} +  \lambda {\bf K}{\bf c} - {\bf K}{\bf x} \right)^{\rm T} 
& = &  0, \nonumber \\
&  & 
\label{eq:final3}
\end{eqnarray}
which is satisfied, iff
\begin{eqnarray}
\left(  {\bf K}{\bf K}{\bf c} +  \lambda {\bf K}{\bf c} - {\bf K}{\bf x} \right) & = & 0.
\label{eq:final4}
\end{eqnarray}
Multiplication with ${\bf K}^{-1}$ from the left, and
rearranging results in Eq.~\ref{eq:l2}.
          
\section{Supplementary Information}
Indices of the 22\,k GDB-8 molecules, to retrieve their geometries from
the 134\,k GDB-9 dataset~\cite{ramakrishnan2014quantum}, 
along with TDDFT, and CC2 excitation energies are collected in 
{\tt gdb8\_22k\_elec\_spec.txt}. 

\bibliography{lit} 

\end{document}